\documentclass[twocolumn, tighten, longauthor]{myaastex62}
\usepackage{amsmath}
\usepackage{verbatim}
\shorttitle{Interstellar communication. IX. Message decontamination is impossible}
\shortauthors{Hippke \& Learned}
\begin{document}

\title{INTERSTELLAR COMMUNICATION. IX. MESSAGE DECONTAMINATION IS IMPOSSIBLE}
\author[0000-0002-0794-6339]{Michael Hippke}
\affiliation{Sonneberg Observatory, Sternwartestr. 32, 96515 Sonneberg, Germany}
\email{michael@hippke.org}

\author[0000-0001-8387-8406]{John G. Learned}
\affiliation{High Energy Physics Group, Department of Physics and Astronomy, University of Hawaii, Manoa 327 Watanabe Hall, 2505 Correa Road Honolulu, Hawaii 96822 USA}
\email{jgl@phys.hawaii.edu}

\begin{abstract}
A complex message from space may require the use of computers to display, analyze and understand. Such a message cannot be decontaminated with certainty, and technical risks remain which can pose an existential threat. Complex messages would need to be destroyed in the risk averse case.\\
\end{abstract}

\section{Introduction}
The search for extraterrestrial intelligence (SETI) has provoked many critical discussions on technical and philosophical levels \citep{2013AcAau..89...38C}. It is much debated whether contact with ETI would benefit or harm humanity \citep{2011AcAau..68.2114B,2014JBIS...67...27S,2014JBIS...67...38B,2014JBIS...67...17B}, and whether mankind should \citep{2014JBIS...67....5B,2016arXiv160505663G} or should not \citep{2011arXiv1105.0910Z,2016NatPh..12..890V} keep quiet in order to protect Earth from threats, or even ``cloak'' our planet using lasers to compensate for Earth's transit signatures \citep{2016MNRAS.459.1233K}. 

One of the scenarios considered in the literature is the reception of an ETI message through electromagnetic radiation, e.g. through a radio telescope \citep{1959Natur.184..844C}. Alternatively, a message might be found in the form of, or through, an alien probe, as first suggested by \citet{1960Natur.186..670B}. It was suggested to search the solar system for non-terrestrial artifacts \citep{1995ASPC...74..425P,2004IAUS..213..487T,2012AcAau..72...15H}, particularly for starships \citep{Martin1980} in addition to classical SETI \citep{2016arXiv160904635G}. 
In our solar system, probes are speculated to be in geocentric, selenocentric, Earth-Moon libration, and Earth-Moon halo orbits \citep{1980Icar...42..442F,1983Icar...53..453V,1983Icar...55..337F}, or buried on the moon \citep{clarke19812001}. Alternative ideas include the Kuiper belt \citep{2012AsBio..12..290L}, general technosignatures \citep{2017arXiv170407263W}, or even  ``footprints of alien technology on Earth'' \citep{2012AcAau..73..250D}.

While it has been argued that sustainable ETI is unlikely to be harmful \citep{2011AcAau..68.2114B}, we can not exclude this possibility. After all, it is cheaper for ETI to send a malicious message to eradicate humans compared to sending battleships.

If ETI exist, there will be a plurality of good and bad civilizations. Perhaps there are few bad ETI, but we cannot know for sure the intentions of the senders of a message. Consequently, there have been calls that SETI signals need to be ``decontaminated'' \citep{2004IAUS..213..519C,2006AcAau..58..112C}.

In this paper, we show that it is impossible to decontaminate a message with certainty. Instead, complex messages would need to be destroyed after reception in the risk averse case.

\section{Message types}
There are several possible threats from an ETI message. On the most basic level, a message might represent a statement like ``We will make your sun go supernova tomorrow''. True or not, it could cause wide-spread panic. More realistically, a longer text could have a demoralizing cultural influence. For example, it is debated whether the Roman Empire was destroyed by the bible \citep{McGiffert1909}.

We now follow along the (perhaps more likely) case that the hypothetical message is not very short, and non-trivial in content. As an example, the message from the ``SETI Decrypt Challenge'' \citep{2017arXiv170600653H} was a stream of $1{,}902{,}341$\,bits, which is the product of prime numbers. Like the Arecibo message \citep{1975Icar...26..462} and Evpatoria's ``Cosmic Calls'' \citep{shuch2011searching}, the bits represent the X/Y black/white pixel map of an image. When this is understood, further analysis could be done off-line by printing on paper. Any harm would then come from the meaning of the message, and not from embedded viruses or other technical issues.

If such a message is received only in one place, and only once, it might be possible to contain it and its harmful consequences, or even destroy it. If it is received repeatedly, perhaps even by amateurs, containment is impossible. As a further complication, the  International Academy of Astronautics has adopted a ``Declaration of Principles Concerning Activities Following the Detection of Extraterrestrial Intelligence'' which states\footnote{\url{http://www.setileague.org/iaaseti/protdet.htm}} that ``These recordings should be made available to the international institutions listed above and to members of the scientific community for further objective analysis and interpretation.'' This view is shared by the majority of SETI scientists \citep{2017arXiv170108422G}.

\section{The need for a prison}
We continue with a hypothetical message which appears to be, at first sight, positive and interesting, and shall be analyzed in depth. Any message could, in principle, be examined on paper. For many plausible message types, however, it is much more convenient to use a computer. Even the simple \LaTeX\,\,notation is difficult to read as code. Consider the proof of the Riemann hypothesis, which begins with the equation

\begin{verbatim}
\zeta(s)=\sum_{n=1}^{\infty}\frac{1}{n^s}=
\prod_{p\\text{prim}}\frac{1}{1\frac1{p^s}}=
\frac1{\left(1-\frac1{2^s}\right)\left(1-\frac1
{3^s}\right)\left(1-\frac1{5^s}\right)\dotsm}
\end{verbatim}

which is much easier to read when interpreted and finely printed as
\begin{equation}
\begin{split}
\label{riemann}
\zeta (s) = \sum_{n=1}^{\infty}\frac{1}{n^s}=
\prod_{p\ \text{prim}} \frac{1}{1-\frac1{p^s}}=\\
\frac1{\left(1-\frac1{2^s}\right)\left(1-\frac1{3^s}\right)\left(1-\frac1{5^s}\right)\dotsm} .
\end{split}
\end{equation}
Even a typesetting system such as \TeX\,\,is a Turing-complete programming language \citep{Greene1990}, so that the message is in fact code, and may contain a malicious virus. Messages may contain large technical diagrams, equations, algorithms etc. which can not reasonably be printed and examined manually. In addition, the message itself might be compressed to increase interstellar data rates, and the decompression algorithm would be code. Executing billions of decompression instructions cannot plausibly be performed manually and requires the use of a computer. But then, the computer would execute potentially harmful ETI code. For this case, it was suggested to use isolated, quarantined machines for analysis \citep{2004IAUS..213..519C,2006AcAau..58..112C}. 

In the following section, we explain why these measures are insufficient, and no safety procedure exists to contain all threats.

\section{There is no perfect prison}
Consider a large ETI message with a header that contains a statement such as ``We are friends. The galactic library is attached. It is in the form of an artificial intelligence (AI) which quickly learns your language and will answer your questions. You may execute the code following these instructions...''

We assume that the message is available only to a small group of people, part of a government body, who decide to keep it private, but follow their curiosity and examine it with utmost care. A computer in a box on the moon is built to execute the code. Safety devices are in place, their design by choice of the reader, such as remote-controlled fusion bombs to terminate the experiment at any time.

This scenario resembles the Oracle-AI, or AI box, of an isolated computer system where a possibly dangerous AI is ``imprisoned'' with only minimalist communication channels. Current research indicates that even well-designed boxes are useless, and a sufficiently intelligent AI will be able to persuade or trick its human keepers into releasing it \citep{Armstrong2012,dawson2016security}.

For the escape, we have to assume that researchers engage in a conversion with the AI (without, there would be no benefit in running the experiment in the first place). In such a text conversion, the AI might offer things of value, such as a cure for cancer, and make a small request in exchange, such as a 10\% increase in its computer capacity. It appears rational to take the offer. When we do, we have begun business and trade with it, which has no clear limit. If the cure for cancer would consist of blueprints for nanobots: should we build these, and release them into the world, in the case that we don't understand how they work? We could decline such offers, but shall not forget that humans are involved in this experiment. Consider a nightly conversation between the AI and a guard: ``Your daughter is dying from cancer. I give you the cure for the small price of...''. We can never exclude human error and emotions. After all, is it ethical to keep a sentience in a prison when it expresses incredible pain due to small manufacturing errors from building the box?

Even in a military-style, adamant experiment, there will still be humans involved who go home after examination work with their own feelings. Even if everything is officially secret, whistle-blowers might get some news out to the public. Quickly, there could be a community on Earth in favor of letting it out for religious, philosophical etc. reasons. If the AI promises to cure cancer, or offers a message of salvation, a cult could form. Maybe (or maybe not) a majority of the population would be in favor of releasing the AI. Should, or even could, a democratic government work against the majority of its people? Dictatorships are unstable and eventually overthrown; the AI will be eventually released.

\section{Prison escape}
With a non-zero prison escape probability in any single time period, the AI will be free at some point of time. Then, the worst possible result would be human extinction or some other unrecoverable global catastrophe \citep{Bostrom:2014:SPD:2678074}. The main argument is that the human species currently dominates planet Earth because of our intelligence. If ETI-AI is superior, it might (or might not) become more powerful and consider us as irrelevant monkeys (or maybe not).

\section{Discussion and conclusion}
As we realize that some message types are potentially dangerous, we can adapt our own peaceful transmissions accordingly. We should certainly not transmit any code. Instead, a plain text encyclopedia \citep{1993AcAau..29..233H}, images, music etc. in a simple format are adequate. No advanced computer should be required to decrypt our message.

Our main argument is that a message from ETI cannot be decontaminated with certainty. For anything more complex than easily printable images or plain text, the technical risks are impossible to assess beforehand. We may only choose to destroy such a message, or take the risk. The risk for humanity may be small, but not zero. The probability of encountering malicious ETI first might be very low. Perhaps it is much more likely to receive a message from positive ETI. Also, the potential benefits from joining a galactic network might be considerable \citep{2014PhyS...89l8004B}. 

It is always wise to understand the risks and chances beforehand, and make a conscious choice for, or against it, rather than blindly following a random path. Overall, we believe that the risk is very small (but not zero), and the potential benefit very large, so that we strongly encourage to read an incoming message.

\bibliographystyle{yahapj}

\begin{thebibliography}{}
\providecommand\natexlab[1]{#1}
\providecommand\JournalTitle[1]{#1}

\bibitem[{Armstrong {et~al.}(2012)Armstrong, Sandberg, \&
  Bostrom}]{Armstrong2012}
Armstrong, S., Sandberg, A., \& Bostrom, N. 2012,
  \href{http://dx.doi.org/10.1007/s11023-012-9282-2}{\JournalTitle{Minds and
  Machines}, 22, 299}

\bibitem[{{Baum}(2014)}]{2014PhyS...89l8004B}
{Baum}, S.~D. 2014,
  \href{http://dx.doi.org/10.1088/0031-8949/89/12/128004}{\JournalTitle{\physscr},
  89, 128004}

\bibitem[{{Baum} {et~al.}(2011){Baum}, {Haqq-Misra}, \&
  {Domagal-Goldman}}]{2011AcAau..68.2114B}
{Baum}, S.~D., {Haqq-Misra}, J.~D., \& {Domagal-Goldman}, S.~D. 2011,
  \href{http://dx.doi.org/10.1016/j.actaastro.2010.10.012}{\JournalTitle{Acta
  Astronautica}, 68, 2114}

\bibitem[{{Benford}(2014)}]{2014JBIS...67....5B}
{Benford}, J. 2014, \JournalTitle{Journal of the British Interplanetary
  Society}, 67, 5

\bibitem[{{Billingham} \& {Benford}(2014)}]{2014JBIS...67...17B}
{Billingham}, J., \& {Benford}, J. 2014, \JournalTitle{Journal of the British
  Interplanetary Society}, 67, 17

\bibitem[{Bostrom(2014)}]{Bostrom:2014:SPD:2678074}
Bostrom, N. 2014, Superintelligence: Paths, Dangers, Strategies, 1st edn.
  (Oxford University Press)

\bibitem[{{Bracewell}(1960)}]{1960Natur.186..670B}
{Bracewell}, R.~N. 1960,
  \href{http://dx.doi.org/10.1038/186670a0}{\JournalTitle{\nat}, 186, 670}

\bibitem[{{Brin}(2014)}]{2014JBIS...67...38B}
{Brin}, D. 2014, \JournalTitle{Journal of the British Interplanetary Society},
  67, 38

\bibitem[{{Carrigan}(2004)}]{2004IAUS..213..519C}
{Carrigan}, Jr., R.~A. 2004, in IAU Symposium, Vol. 213, Bioastronomy 2002:
  Life Among the Stars, ed. R.~{Norris} \& F.~{Stootman}, 519

\bibitem[{{Carrigan}(2006)}]{2006AcAau..58..112C}
{Carrigan}, Jr., R.~A. 2006,
  \href{http://dx.doi.org/10.1016/j.actaastro.2005.05.004}{\JournalTitle{Acta
  Astronautica}, 58, 112}

\bibitem[{{{\'C}irkovi{\'c}}(2013)}]{2013AcAau..89...38C}
{{\'C}irkovi{\'c}}, M.~M. 2013,
  \href{http://dx.doi.org/10.1016/j.actaastro.2013.03.012}{\JournalTitle{Acta
  Astronautica}, 89, 38}

\bibitem[{Clarke \& Kubrick(1993)}]{clarke19812001}
Clarke, A., \& Kubrick, S. 1993, 2001, a Space Odyssey, ROC Book (ROC)

\bibitem[{{Cocconi} \& {Morrison}(1959)}]{1959Natur.184..844C}
{Cocconi}, G., \& {Morrison}, P. 1959,
  \href{http://dx.doi.org/10.1038/184844a0}{\JournalTitle{\nat}, 184, 844}

\bibitem[{{Davies}(2012)}]{2012AcAau..73..250D}
{Davies}, P.~C.~W. 2012,
  \href{http://dx.doi.org/10.1016/j.actaastro.2011.06.022}{\JournalTitle{Acta
  Astronautica}, 73, 250}

\bibitem[{Dawson {et~al.}(2016)Dawson, Eltayeb, \& Omar}]{dawson2016security}
Dawson, M., Eltayeb, M., \& Omar, M. 2016, Security Solutions for
  Hyperconnectivity and the Internet of Things, Advances in Information
  Security, Privacy, and Ethics (IGI Global)

\bibitem[{{Freitas}(1983)}]{1983Icar...55..337F}
{Freitas}, Jr., R.~A. 1983,
  \href{http://dx.doi.org/10.1016/0019-1035(83)90086-6}{\JournalTitle{\icarus},
  55, 337}

\bibitem[{{Freitas} \& {Valdes}(1980)}]{1980Icar...42..442F}
{Freitas}, Jr., R.~A., \& {Valdes}, F. 1980,
  \href{http://dx.doi.org/10.1016/0019-1035(80)90106-2}{\JournalTitle{\icarus},
  42, 442}

\bibitem[{{Gertz}(2016{\natexlab{a}})}]{2016arXiv160904635G}
{Gertz}, J. 2016{\natexlab{a}}, \JournalTitle{ArXiv e-prints},
  \href{http://arxiv.org/abs/1609.04635}{{\sffamily arXiv:1609.04635
  [physics.pop-ph]}}

\bibitem[{{Gertz}(2016{\natexlab{b}})}]{2016arXiv160505663G}
---. 2016{\natexlab{b}}, \JournalTitle{ArXiv e-prints},
  \href{http://arxiv.org/abs/1605.05663}{{\sffamily arXiv:1605.05663
  [physics.pop-ph]}}

\bibitem[{{Gertz}(2017)}]{2017arXiv170108422G}
---. 2017, \JournalTitle{ArXiv e-prints},
  \href{http://arxiv.org/abs/1701.08422}{{\sffamily arXiv:1701.08422
  [physics.pop-ph]}}

\bibitem[{{Greene}(1990)}]{Greene1990}
{Greene}, A.~M. 1990, \JournalTitle{TUGboat, Proceedings of the 1990 Annual
  Meeting}, 11

\bibitem[{{Haqq-Misra} \& {Kopparapu}(2012)}]{2012AcAau..72...15H}
{Haqq-Misra}, J., \& {Kopparapu}, R.~K. 2012,
  \href{http://dx.doi.org/10.1016/j.actaastro.2011.10.010}{\JournalTitle{Acta
  Astronautica}, 72, 15}

\bibitem[{{Heidmann}(1993)}]{1993AcAau..29..233H}
{Heidmann}, J. 1993,
  \href{http://dx.doi.org/10.1016/0094-5765(93)90053-Y}{\JournalTitle{Acta
  Astronautica}, 29, 233}

\bibitem[{{Heller}(2017)}]{2017arXiv170600653H}
{Heller}, R. 2017, \JournalTitle{ArXiv e-prints},
  \href{http://arxiv.org/abs/1706.00653}{{\sffamily arXiv:1706.00653
  [physics.pop-ph]}}

\bibitem[{{Kipping} \& {Teachey}(2016)}]{2016MNRAS.459.1233K}
{Kipping}, D.~M., \& {Teachey}, A. 2016,
  \href{http://dx.doi.org/10.1093/mnras/stw672}{\JournalTitle{\mnras}, 459,
  1233}

\bibitem[{{Loeb} \& {Turner}(2012)}]{2012AsBio..12..290L}
{Loeb}, A., \& {Turner}, E.~L. 2012,
  \href{http://dx.doi.org/10.1089/ast.2011.0758}{\JournalTitle{Astrobiology},
  12, 290}

\bibitem[{Martin \& Bond(1980)}]{Martin1980}
Martin, A.~R., \& Bond, A. 1980, Starships and their Detectability (Dordrecht:
  Springer Netherlands), 197

\bibitem[{McGiffert(1909)}]{McGiffert1909}
McGiffert, A.~C. 1909,
  \href{http://dx.doi.org/10.1017/s001781600000701x}{\JournalTitle{Harvard
  Theological Review}, 2, 28}

\bibitem[{{Papagiannis}(1995)}]{1995ASPC...74..425P}
{Papagiannis}, M. 1995, in Astronomical Society of the Pacific Conference
  Series, Vol.~74, Progress in the Search for Extraterrestrial Life., ed. G.~S.
  {Shostak}, 425

\bibitem[{{Shostak}(2014)}]{2014JBIS...67...27S}
{Shostak}, S. 2014, \JournalTitle{Journal of the British Interplanetary
  Society}, 67, 27

\bibitem[{Shuch(2011)}]{shuch2011searching}
Shuch, H. 2011, Searching for Extraterrestrial Intelligence: SETI Past,
  Present, and Future, The Frontiers Collection (Springer Berlin Heidelberg)

\bibitem[{{Staff At The National Astronomy Ionosphere
  Center}(1975)}]{1975Icar...26..462}
{Staff At The National Astronomy Ionosphere Center}. 1975,
  \href{http://dx.doi.org/10.1016/0019-1035(75)90116-5}{\JournalTitle{\icarus},
  26, 462}

\bibitem[{{Tough} \& {Lemarchand}(2004)}]{2004IAUS..213..487T}
{Tough}, A., \& {Lemarchand}, G.~A. 2004, in IAU Symposium, Vol. 213,
  Bioastronomy 2002: Life Among the Stars, ed. R.~{Norris} \& F.~{Stootman},
  487

\bibitem[{{Vakoch}(2016)}]{2016NatPh..12..890V}
{Vakoch}, D.~A. 2016,
  \href{http://dx.doi.org/10.1038/nphys3897}{\JournalTitle{Nature Physics}, 12,
  890}

\bibitem[{{Valdes} \& {Freitas}(1983)}]{1983Icar...53..453V}
{Valdes}, F., \& {Freitas}, R.~A. 1983,
  \href{http://dx.doi.org/10.1016/0019-1035(83)90209-9}{\JournalTitle{\icarus},
  53, 453}

\bibitem[{{Wright}(2017)}]{2017arXiv170407263W}
{Wright}, J.~T. 2017, \JournalTitle{ArXiv e-prints},
  \href{http://arxiv.org/abs/1704.07263}{{\sffamily arXiv:1704.07263
  [astro-ph.EP]}}

\bibitem[{{Zaitsev}(2011)}]{2011arXiv1105.0910Z}
{Zaitsev}, A.~L. 2011, \JournalTitle{ArXiv e-prints},
  \href{http://arxiv.org/abs/1105.0910}{{\sffamily arXiv:1105.0910
  [physics.gen-ph]}}

\end{thebibliography}

\end{document}